\documentstyle[prl,aps]{revtex}
\begin{document}
\def\be{\begin{equation}}
\def\ee{\end{equation}}
\def\bea{\begin{eqnarray}}
\def\eea{\end{eqnarray}}
\title{Quasi-long range order in the random anisotropy Heisenberg
model}
\author{D.E. Feldman}
\address{ Landau Institute for Theoretical Physics,
 142432, Chernogolovka, Moscow region, Russia }
\maketitle

\begin{abstract}
The large distance behaviors of the random field and random anisotropy
Heisenberg models are studied with the functional renormalization
group in $4-\epsilon$ dimensions. The random anisotropy model is
found to have a phase with the infinite correlation radius at low
temperatures and weak disorder. The correlation function of the
magnetization obeys a power law $\langle {\bf m}({\bf r}_1) {\bf m}({\bf
r}_2)\rangle\sim| {\bf r}_1-{\bf r}_2|^{-0.62\epsilon}$. The magnetic
susceptibility diverges at low fields as $\chi\sim H^{-1+0.15\epsilon}$.
In the random field model the correlation radius is found to be finite
at the arbitrarily weak disorder.
\end{abstract}

The effect of impurities on the order in condensed matter is
interesting since the disorder is almost inevitably present in any
system. If the disorder is weak, the short range order is the same as
in the pure system. However, the large distance behavior can be
strongly modified by the arbitrarily weak disorder. This happens in
the systems of continuous symmetry in presence of the random
symmetry breaking field \cite{IM}. The first experimental example of
this kind is the amorphous magnet \cite{HPZ,AM}. During the last
decade a lot of other related objects were found. These are liquid
crystals in the porous media \cite{LQ}, nematic elastomers \cite{NE},
He-3 in aerogel \cite{He3} and vortex phases of
impure superconductors \cite{HTSC}. The
nature of the low-temperature phases of these systems is still
unclear. The only reliable statement is that a long range
order is absent \cite{IM,L,P,AW}. However, other details
of the large distance behavior are poorly understood.

The neutron scattering \cite{Xray} reveals sharp Bragg peaks in
impure superconductors at
low temperatures and weak external magnetic fields. Since the vortices
can not form a regular lattice \cite{L}, it is tempting to assume,
that there is a quasi-long range order (QLRO), that is the correlation
radius is infinite and correlation functions depend on the distance
slow. Recent theoretical \cite{RFXY} and numerical \cite{numXY}
studies of the random field XY model, which is the simplest model of
the vortex system in the impure superconductor \cite{HTSC}, support
this picture. The theoretical advances \cite{RFXY} are afforded by two
new technical approaches: the functional renormalization group
\cite{FRG} and the replica variational method \cite{MP}. These
methods are free from drawbacks of the standard renormalization group
and give reasonable results. The variational method regards a
possibility of spontaneous replica symmetry breaking and treats the
fluctuations approximately. On the other hand, the functional
renormalization group provides a subtle analysis of the fluctuations
about the replica symmetrical ground state.  Surprisingly, the methods
suggest close and sometimes even the same results.

Both techniques were originally suggested for the random
manifolds \cite{FRG,MP} and then allowed to obtain information about
some other disordered systems with the abelian symmetry
\cite{RFXY,F,EN,RT}.
It is less known about the non-abelian systems. The simplest of
them are the random field \cite{IM} and random anisotropy \cite{HPZ}
Heisenberg models. The latter was introduced as a model of the
amorphous magnet \cite{HPZ}. In spite of a long discussion, the
question about QLRO in these models is still open. There is an
experimental evidence in favor of no QLRO  \cite{B}.
On the other hand, recent numerical simulations \cite{num} support the
possibility of QLRO in these systems. The only theoretical approach,
developed up to now, is based on the spherical approximation
\cite{spher}.  However, there is no reason for this approximation to
be valid.

In this letter we study the random field and random anisotropy
Heisenberg models in $4-\epsilon$ dimensions with the functional
renormalization group. The large distance behaviors of the systems are
found to be quite different.  While in the random field model the
correlation radius is always finite, the random anisotropy Heisenberg
model has a phase with QLRO. In this phase the correlation function of
the magnetization obeys a power law and the magnetic susceptibility
diverges at low fields.

To describe the large distance behavior at low temperatures we use the
classical nonlinear $\sigma$-model with the Hamiltonian

\be
\label{1}
H=\int d^D x[J\sum_\mu\partial_\mu{\bf n}({\bf x})
\partial_\mu{\bf n}({\bf x}) + V_{\rm imp}({\bf x})],
\ee
where ${\bf n}({\bf x})$ is the unit vector of the magnetization,
$V_{\rm imp}({\bf x})$ the random potential. In the random field case
it has the form

\be
\label{2}
V_{\rm imp}=-\sum_\alpha h_\alpha({\bf x})n_\alpha({\bf x});
\alpha=x,y,z,
\ee
where the random field ${\bf h}({\bf x})$ has a Gaussian
distribution and $\langle h_\alpha({\bf x})h_\beta({\bf
x}')\rangle=A^2\delta({\bf x}-{\bf x}')\delta_{\alpha\beta}$. In the
random anisotropy case the random potential is given by the equation

\be
\label{3}
V_{\rm imp}=-\sum_{\alpha,\beta}\tau_{\alpha \beta}({\bf
x})n_\alpha({\bf x})n_\beta({\bf x}); \alpha,\beta=x,y,z,
\ee
where $\tau_{\alpha\beta}({\bf x})$ is a Gaussian random variable,
$\langle\tau_{\alpha\beta}({\bf x})\tau_{\gamma\delta}({\bf
x}')\rangle=A^2\delta_{\alpha\gamma}\delta_{\beta\delta}\delta({\bf
x}-{\bf x}')$. Random potential (\ref{3}) corresponds to the
same symmetry as a more conventional choice
$V_{\rm imp}=-({\bf hn})^2$, but is more convenient for the further
discussion.

The Imry-Ma argument \cite{IM,P} suggests that in our problem the long
range order is absent at any dimension $D<4$. One can estimate the
Larkin length, up to which there are strong ferromagnetic
correlations, with the following qualitative renormalization group
(RG) approach. Let one remove the fast modes and rewrite the
Hamiltonian in terms of the block spins, corresponding to the scale
$L=ba$, where $a$ is the ultraviolet cut-off. Then let one make
rescaling so that the Hamiltonian would restore its initial form with
new constants $A(L), J(L)$. Dimensional analysis provides estimations

\be
\label{4}
J(L)\sim b^{D-2} J(a); A(L)\sim b^{D/2}A(a)
\ee
To estimate the typical angle $\phi$ between neighbor block spins, one
notes that the effective field, acting on each spin, has two
contributions: the exchange contribution and the random one. The
exchange contribution of order $J(L)$ is oriented along the local
average direction of the magnetization. The random contribution of
order $A(L)$ may have any direction. This allows one to write at low
temperatures that $\phi(L)\sim A(L)/J(L)$. The Larkin length
corresponds to the condition $\phi(L)\sim 1$ and equals $L\sim
(J/A)^{2/(4-D)}$ in agreement with the Imry-Ma argument \cite{IM}.
If Eq. (\ref{4}) were exact, the Larkin length could be interpreted as
the correlation radius. However, there are two sources of corrections
to Eq. (\ref{4}). Both of them are relevant already at the derivation
of the RG equation for the pure system in $2+\epsilon$ dimensions
\cite{Pol}. The first source is the renormalization due to the
interaction and the second one results from the rescaling of the
magnetization, which is necessary to ensure the fixed length condition
${\bf n}^2=1$. The leading corrections to Eq. (\ref{4}) are
proportional to $\phi^2 J, \phi^2 A$. Thus, the RG equation for the
combination $(A(L)/J(L))^2$ is the following

\be
\label{6}
\frac{d}{d \ln L}\left(\frac{A(L)}{J(L)}\right)^2=
\epsilon\left(\frac{A(L)}{J(L)}\right)^2+
c\left(\frac{A(L)}{J(L)}\right)^4, \epsilon=4-D
\ee
If the constant $c$ in Eq. (\ref{6}) is positive, the Larkin length is
the correlation radius indeed. But if $c<0$, the RG equation has a fixed
point, corresponding to the phase with the infinite correlation
radius. As it is seen below, both situations are possible, depending
on the system.

To derive the RG equations in a systematic way we use the method,
suggested by Polyakov \cite{Pol} for the pure system. The same
consideration as in the XY \cite{RFXY} and random manifold \cite{FRG}
models suggests that near a zero-temperature fixed point in
$4-\epsilon$ dimensions there is an infinite set of relevant operators.
After replica averaging, the relevant part of the effective replica
Hamiltonian can be represented in the following form

\be
\label{7}
H_R=\int d^D x[\sum_a\frac{1}{2T}\sum_\mu
\partial_\mu{\bf n}_a\partial_\mu{\bf n}_a - \sum_{ab}\frac{R({\bf
n}_a{\bf n}_b)}{T^2}],
\ee
where $a,b$ are replica indices, $R(z)$ is some function, $T$ the
temperature. In the random anisotropy case the function $R(z)$ is even
due to the symmetry with respect to changing the sign of the
magnetization.

The one-loop RG equations in $4-\epsilon$ dimensions are obtained by
a straightforward combination of the methods of Refs. \cite{FRG} and
\cite{Pol}. The equations below are given for the arbitrary number
$N$ of
the components of the magnetization. In the Heisenberg model $N=3$.
The RG equations become simpler after substitution for the argument of
the function $R(z)$: $z=\cos\phi$. In terms of this new variable one
has to find even periodic solutions $R(\phi)$. The period is $2\pi$
in the random field case and $\pi$ in the random anisotropy case. In
a zero-temperature fixed point the one-loop equations are

\be
\label{8}
\frac{d\ln T}{d\ln L}= -(D-2) - 2(N-2)R''(0)+O(R^2,T);
\ee
\begin{eqnarray}
0=\frac{dR(\phi)}{d \ln L}=\epsilon R(\phi) + (R''(\phi))^2 - 2R''(\phi)
R''(0) -  & & \nonumber\\
\label{9}
(N-2)[4R(\phi)R''(0)+2{\rm ctg}\phi R'(\phi) R''(0) -
\left(\frac{R'(\phi)}{\sin\phi}\right)^2] + O(R^3,T) & &
\end{eqnarray}
The two-spin correlation function is given by the expression
$\langle{\bf n}({\bf x}){\bf n}({\bf x}')\rangle\sim|{\bf x}-{\bf
x}'|^{-\eta}$, where

\be
\label{11}
\eta=-2(N-1)R''(0)
\ee
The same equations (\ref{8}-\ref{11}) were derived by a different
method in Ref. \cite{DF}. In that paper the critical behavior in
$4+|\epsilon|$ dimensions was studied by considering analytical
fixed point solutions $R(\phi)$. In the Heisenberg model, analytical
solutions are absent and they are unphysical for $N\ne 3$ \cite{DF}.
In this letter we search for non-analytical $R(\phi)$.

As shown below, the random field model can be completely studied
by analytical means. In the random anisotropy case, one has to solve
Eq. (\ref{9}) numerically. Since coefficients of Eq. (\ref{9}) are
large as $\phi\rightarrow 0$, it is convenient to use the expansion of
$R(\phi)$ over $|\phi|$ at small $\phi$.  At larger $\phi$ the
equation is integrated by the Runge-Kutta method. The solutions to be
found have zero derivatives at $\phi=0,\pi/2$. At $N=3$ the solution
with largest $|R''(0)|$, which corresponds to $\eta=0.62\epsilon$
(\ref{11}), has two zeroes in the interval $[0,\pi]$. There are also
solutions with 4 and more zeroes. They all correspond to
$\eta<0.5\epsilon$. We shall see below, that these solutions are
unstable.

To test the stability of the solution with two zeroes, we use an
approximate method. The instability to the constant shift
of the function $R(\phi)$ has no interest for us, since the constant
shifts do not change the correlators \cite{FRG}. To study the stability
to the other perturbations, it is convenient to rewrite Eq.
(\ref{9}), substituting $\omega(R''(\phi))^2$ for $(R''(\phi))^2$.
The case of interest is $\omega=1$, but at $\omega=0$ the equation can
be solved exactly. The solution at $\omega=1$ can be found with the
perturbation theory over $\omega$. The exact solution at $\omega=0$ is
$R_{\omega=0}(\phi)=\epsilon(\cos\phi/24+1/120)$. The perturbative
expansion provides the following asymptotic series for $\eta$:
$\eta=\epsilon(0.67-0.08\omega+0.14\omega^2-\dots)$. The resulting
estimation $\eta=\epsilon(0.67\pm0.08)$ agrees with the numerical
result well. This allows us to expect that the stability analysis of
the solution $R_{\omega=0}$ of the equation with $\omega=0$ provides
information about the stability of the solution of Eq. (\ref{9}).
A simple calculation shows that $R_{\omega=0}$ is stable in the linear
approximation. Thus, there is a stable zero-temperature fixed point of
the RG equations with the critical exponent of the correlation
function

\be
\label{14}
\eta=0.62\epsilon
\ee
The critical exponent $\gamma$ of the magnetic susceptibility
$\chi(H)\sim H^{-\gamma}$ in the weak uniform field $H$ is given
by the equation

\be
\label{15}
\gamma=1+(N-1)R''(0)/2=1-0.15\epsilon
\ee

Let us demonstrate the absence of physically acceptable fixed points
in the random field case. We derive some inequality for critical
exponents. Then we show that the inequality has no solutions.
We use the rigorous inequality for the connected and
disconnected correlation functions \cite{SwSo}

\be
\label{17}
\langle{\bf n}_a({\bf q}){\bf n}_a(-{\bf q})\rangle -
\langle{\bf n}_a({\bf q}){\bf n}_b(-{\bf q})\rangle \le
{\rm const}\sqrt{\langle{\bf n}_a({\bf q}){\bf n}_a(-{\bf q})\rangle},
\ee
where ${\bf n}({\bf q})$ is a Fourier-component of the magnetization,
$a,b$ are replica indices. In a fixed point, Eq. (\ref{17}) provides
an inequality for the critical exponents of the connected and
disconnected correlation functions \cite{SwSo}. The large distance
behavior of the connected correlation function in a zero-temperature
fixed point can be derived from the expression
$\chi\sim\int\langle\langle {\bf n}({\bf 0}){\bf n}({\bf
x})\rangle\rangle d^D x$ and the critical exponent of the
susceptibility (\ref{15}). Finally, one obtains the following
relation

\be
\label{18}
4-D \le\frac{3-N}{N-1}\eta,
\ee
where $\eta$ is given by Eq. (\ref{11}). This equation does not have
solutions at $N=3$. At $N>3$ Eq. (\ref{18}) is incompatible with the
requirement $\eta>0$. Thus, there are no accessible fixed points for
$N\ge 3$.

In the previous paragraph, inequality (\ref{18}) is derived for the
model (\ref{1}) with the Gaussian random field (\ref{2}). It can also
be extended to a more general situation (\ref{7}). If one adds a weak
Gaussian random field (\ref{2}) to any Hamiltonian, it suffices for
Eq. (\ref{17}) to become valid. The addition of the Gaussian random
field corresponds to the transformation $R({\bf n}_a{\bf
n}_b)\rightarrow R({\bf n}_a{\bf n}_b)+\Delta{\bf n}_a{\bf n}_b$ in Eq.
(\ref{7}), where $\Delta>0$ is a constant.
Thus, if at some $\Delta$ the function $\tilde
R({\bf n}_a{\bf n}_b) = R({\bf n}_a{\bf n}_b) - \Delta{\bf n}_a{\bf
n}_b$ is possible as a disorder-induced term in Eq. (\ref{7}), 
then Eq. (\ref{17}) is valid for the
system with the disorder-induced term $R({\bf n}_a{\bf n}_b)$. Finally,
we conclude, that inequality (\ref{18}) may be broken only for
Hamiltonians (\ref{7}), which lie outside the physically acceptable
region or on its border. This suggests the strong coupling regime with a
presumably finite correlation radius.

In the random anisotropy case a similar consideration uses the
connected and disconnected correlation functions of the field
$(n_x({\bf r})n_y({\bf r}))$ in presence of Gaussian disorder
(\ref{3}). The resulting condition for the critical exponent, $\eta\ge
(N-1)\epsilon/4$, rules out all but one fixed points of RG
equation (\ref{9}).

The question of the large distance behavior of the random field and
random anisotropy Heisenberg models was discussed by Aharony and Pytte
on the basis of an approximate equation of state \cite{AP}. They also
obtained QLRO in the random anisotropy case and its absence in the
random field model. However, we believe, that this is an occasional
coincidence, since the equation of state \cite{AP} is valid only in
the first order in the strength of the disorder, while higher orders
are crucial for critical properties \cite{G}. In particular, the
approach \cite{AP} incorrectly predicts the absence of QLRO in the
random field XY model and its presence in the random anisotropy
spherical model. It also provides incorrect critical exponents in the
Heisenberg case.

The random anisotropy Heisenberg model is relevant for the amorphous
magnets \cite{HPZ}. In the same time, for their large distance
behavior the dipole interaction may be important \cite{B}. Besides, a
weak nonrandom anisotropy is inevitably present due to mechanical
stresses.

In conclusion, we have found, that the random anisotropy Heisenberg
model has the infinite correlation radius and a power dependence of
the correlation function of the magnetization on the distance at low
temperatures and weak disorder in $4-\epsilon$ dimensions. On the
other hand, the correlation radius of the random field Heisenberg model
is always finite.

%\acknowledgments

The author is thankful to E. Domany, G. Falkovich, Y. Gefen,
S.E. Kor\-shu\-nov, Y.B. Levinson, V.L. Pokrovskiy and A.V. Shytov for
useful discussions. This work was supported by RFBR grant 96-02-18985
and by grant 96-15-96756 of the Russian Program of Leading Scientific
Schools.

\end{document}